\documentclass[conference]{IEEEtran}
\usepackage{blindtext, graphicx}
\usepackage[caption=false]{subfig}
\begin{document}

\title{Evaluating Resilience of Encrypted Traffic Classification against Adversarial Evasion Attacks}

\author{\IEEEauthorblockN{Ramy Maarouf, Danish Sattar, and Ashraf Matrawy}
\IEEEauthorblockA{School of Information Technology\\
Carleton University, Ottawa, Canada\\
\textit {Email: [Ramy.Maarouf, Danish.Sattar, Ashraf.Matrawy]@carleton.ca}}}

\maketitle

\begin{abstract}
Machine learning and deep learning algorithms can be used to classify encrypted Internet traffic. Classification of encrypted traffic can become more challenging in the presence of adversarial attacks that target the learning algorithms. In this paper, we focus on investigating the effectiveness of different evasion attacks and see how resilient machine and deep learning algorithms are. Namely, we test C4.5 Decision Tree, K-Nearest Neighbor (KNN), Artificial Neural Network (ANN), Convolutional Neural Networks (CNN) and Recurrent Neural Networks (RNN). In most of our experimental results, deep learning shows better resilience against the adversarial samples in comparison to machine learning. Whereas, the impact of the attack varies depending on the type of attack.
\end{abstract}


\IEEEpeerreviewmaketitle

\section{Introduction}
The amount of Internet traffic is snowballing. One of the key issues for companies and Internet Service Providers (ISPs) is the ability to precisely identify the applications flowing in their networks. Network traffic classification is a key tool to monitor the network activity and detect threats in a specific network \cite{dainotti2012issues}. 

The growing trend of traffic encryption and protocol encapsulation in the network poses challenges, which make it harder to identify or classify the traffic accordingly \cite{xue2013traffic}. Artificial Intelligence (AI) techniques, such as Machine Learning and Deep Learning, can be applied to detect and characterize the type of encrypted traffic in the network. The concept of deep learning consists of creating hierarchical representations that are complex that involve the creation of simple building blocks to solving high-level problems \cite{vigneswaran2018evaluating}. To the best of our knowledge, this is the first research that elaborates a comprehensive comparison between deep learning and machine learning in classifying encrypted multilabel traffic applications at adversarial-free and adversarial attack environments.

This paper analyzes the resilience of machine and deep learning techniques against adversarial samples in encrypted traffic classification. The experiments were evaluated with the average metrics of Precision and Recall to calculate the F1-Score. The main contributions of this paper are summarized as follows:

\begin{itemize}
\item We evaluate the performance of five machine and deep learning algorithms in the adversarial-free environment i.e., C4.5, KNN, ANN, CNN, and RNN on two benchmark datasets: ISCX VPN-NonVPN \cite{draper2016characterization} and NIMS \cite{alshammari2011can} \cite{alshammari2008investigating} \cite{alshammari2007flow} using the same feature selection method: Mutual Information.
\item We test and evaluate the resilience of encrypted traffic classification using machine learning and deep learning algorithms against Zeroth Order Optimization (ZOO) \cite{chen2017zoo}, Projected Gradient Descent (PGD) \cite{madry2018towards}, and DeepFool \cite{moosavi2016deepfool} evasion adversarial attacks.
\item We compare the performance of deep learning and machine learning in an adversarial-free and adversarial attack environment for each dataset, respectively.
\end{itemize}

The remainder of this paper is organized as follows. In section II, we review the related work in the field of network traffic classification. A brief background about machine learning and deep learning algorithms, followed by a detailed overview of adversarial attack methods in section III. Experimental setup and results will be presented in sections IV and V. Finally, the conclusions are discussed in section VI.

\section{Related Work}
This section discusses the state-of-the-art work in traffic classification using machine learning and deep learning algorithms and various studies of adversarial attacks.

Vigneswaran et al. \cite{vigneswaran2018evaluating} presented a comparison of shallow and deep neural networks for intrusion detection systems. They trained deep learning and classic machine learning algorithms using the same dataset. Their results showed that the accuracy, precision, and recall performance rates of deep learning are better than all classic machine learning algorithms.

Taher et al. \cite{taher2019network} proposed a supervised machine learning system for network traffic classification. They trained one machine learning, Support Vector Machine (SVM), and one deep learning, Artificial Neural Network (ANN) algorithm with the same NSL-KDD dataset and applied feature selection methods. They wanted to detect whether the traffic was malicious or benign. They found that the ANN with feature selection performed better than the SVM.

Wang et al. \cite{wang2017end} introduced a traffic classification method with one-dimensional convolution neural networks. Their approach was unique since they proposed an end-to-end framework that learns nonlinear relationships between raw input and the expected output. They built a model that could classify traffic on various levels using feature extraction and feature selection. Lotfollahi et al. \cite{lotfollahi2020deep} presented a framework that can handle both traffic characterization and application identification. This scheme is based on a Convolutional Neural Network (CNN) and can reduce feature engineering to achieve high accuracy.

Xu et al. \cite{xu2020adversarial} defined each type of adversarial attack and divided them into three categories: gradient-masking/obfuscation, robust optimization, and adversarial example detection. They focused on the adversarial examples of deep learning models and evaluated most research efforts on attacking various deep neural networks in different applications.

In Wang et al. \cite{wang2018deep} research, the authors implemented several adversarial attack algorithms such as JSMA, FGSM, Deepfool, and Carlini Wagner (CW) attacks \cite{carlini2017towards}. They analyzed the performances of state-of-the-art attack algorithms against deep learning algorithms using the NSL-KDD dataset. They showed that neural network models can be fooled using adversarial examples even if different datasets were used and trained. They also proved that generating adversarial examples can be affected by significant features regarding their rates to be perturbed by an adversary.

\section{Background}

\subsection{Machine Learning}
Machine Learning can be defined as an application of AI that gives computers the power to extract knowledge automatically and enhance their performance from example data or experience without any exploit programming \cite{singh2016review}. It is a set of algorithms that can define data, learn from it, and then apply this gained knowledge to form smart decisions.

According to the business problem being addressed many techniques are derived from the type and volume of data. One of them is supervised learning, which begins with an established set of data and a certain understanding of how the data is classified. It is responsible for searching for patterns in data with labeled features that can be applied to an analytic process \cite{singh2016review}.

In our work, we will use two well-known supervised machine learning classifiers, the C4.5 Decision Tree and the K-Nearest Neighbor (KNN) algorithms.

\begin{itemize}
\item \textbf{Decision Tree (C4.5)}: It generates a decision tree for classification. It is often referred to as a statistical classifier. It can accept data with categorical or numerical values as it uses information gain as splitting criteria \cite{sharma2016survey}.

\item \textbf{K-Nearest Neighbor (KNN)}: It is an algorithm based on a distance function that measures the difference or similarity between two instances \cite{jiang2007survey}.
\end{itemize}

\subsection{Deep Learning}
Deep Learning is one of the advanced and robust machine learning strategies these days. Deep learning has naturally emerged as researchers began modelling complex problems in domains such as computer vision and voice recognition \cite{kusiak2020convolutional}. Its models learn by extract features from a large labeled dataset with high efficiency and accuracy \cite{papernot2016distillation}. Unlike machine learning, it defeats the need for domain expertise as it works on learning high-level features from data in an incremental manner. Deep learning strength depends on some criteria, starts with; How complex is the problem? What type of prediction is needed? How important is the accuracy to work vs. the interpretability? Will there be enough quality labeled data? How many resources will allocate for work? \cite{8397411}

Most deep learning models are built on neural networks and divided into three linear models of different layers \cite{vinayakumar2017applying}. The first layer is the input layer that provides the information from the outside world and passes on this information to the next layer called the hidden layer. This layer performs the needed computations and transfers the generated information to the output layer responsible for showing the predicted results. The neural network structure is the same as a human brain, formed of cells called neurons, responsible for transmitting electrical signals to each other based on the received signals from other neurons \cite{uhrig1995introduction}.

We will perform our work using Artificial Neural Network (ANN), Convolutional Neural Network (CNN), and Recurrent Neural Networks (RNN).

\begin{itemize}
\item \textbf{Artificial Neural Network (ANN)}: It is a strategy created to simulate the human brain for pattern recognition. It consists of three layers, one layer for input, one for output, and at least one layer hidden between them \cite{abiodun2018state}. The ANN architecture is constructed with two hidden layers of 180 neurons in layer 1 and 160 neurons in layer 2.

\item \textbf{Convolutional Neural Network (CNN)}: It is a technique inspired by the visual mechanisms of living organs and used with dimensional inputs. CNNs reduce the number of learning parameters significantly by using a set of convolution filters (kernels) in convolutional layers \cite{khan2020survey}. The CNN architecture is established with two hidden layers of 84 neurons in layer 1 and 64 neurons in layer 2.

\item \textbf{Recurrent Neural Network (RNN)}: It is an algorithm designed for sequential data, and its learning methods differs from other deep learning model \cite{mikolov2010recurrent}. The RNN architecture is built with two hidden layers of 84 neurons in layer 1 and 64 neurons in layer 2.
\end{itemize}

\subsection{Adversarial Attacks}
In a classification problem, machine learning and deep learning algorithms aim to classify input data into one of each class. Since training the models depends on data, classification function can be manipulated by perturbations input called adversarial samples, which can mislead any model towards inaccurate classification \cite{45818} \cite{43405}. Adversarial attacks are the method of adding noise to the original sample in such a way as to make the decision boundary between the regular and the crafted data shift and inaccurate.

We utilize three attack methods from the literature for machine learning and deep learning models to generate adversarial samples. We use Zeroth Order Optimization (ZOO) \cite{chen2017zoo}, which is used to directly estimate the gradients of the targeted neural networks from function values and apply a first-order optimization algorithm for generating adversarial examples \cite{Golovin2020Gradientless}. However, Projected Gradient Descent (PGD) \cite{madry2018towards} attempts to find the perturbation that maximizes the loss of a model on a particular input while keeping the size of the perturbation smaller than a specified amount. Since it is a multi-step with a negative loss function, it overcomes the network overfit problem and became more robust than the Fast Gradient Sign Method (FGSM) adversarial samples, which come from it. The DeepFool attack \cite{moosavi2016deepfool} is an algorithm that can compute adversarial examples and defeat state-of-art classifiers to change output prediction labels. It searches for the minimal perturbation to fool a classifier and uses concepts from geometry to direct the search. DeepFool attack is considered an efficient method to evaluate the robustness of classifiers.

\section{Experimental Setup}
Machine learning models were implemented using Scikit-learn, and deep learning models were executed using TensorFlow and Keras \cite{brownlee2016deep}. The experiments were performed on a 64-bit Windows machine with 16 GB memory and eight Intel core processors 3.00 GHz. We used the Scientific Python Development Environment (Spyder) \cite{raybaut2009spyder}, included with the Anaconda platform, to write and run our code from there. We used the open-source IBM Robustness Toolbox (ART) framework for generating the adversarial samples \cite{nicolae2018adversarial}. ART is a Python library that provides tools to build defenses against several machine learning models and test them against adversarial threats.

In our paper, we assume our adversarial attacks are white-box evasion attacks. They attack the mentioned algorithms during the test time to misclassify the traffic samples. We conducted two sets of experiments to evaluate the resilience of each machine learning and deep learning algorithm and the resilience against adversarial evasion attacks. The details of each experiment are described in sections B and C.

\subsection{Datasets}
Since the performance of machine learning and deep learning algorithms depends on the dataset to some extent, classifier accuracy differs between various datasets. In our research, we use two encrypted traffic classification public datasets: ISCX VPN-NonVPN and NIMS.

ISCX VPN-NonVPN traffic Dataset \cite{draper2016characterization} is a captured network traffic set consisting of different types of traffic and applications. There are two data formats in this traffic dataset, labeled flow features (ARFF format) and unlabeled raw traffic (PCAP format). It contains packets captured over Virtual Private Network (VPN) sessions and regular Non-VPN encrypted sessions. The dataset objective was to identify different types of applications using VPN and Non-VPN sessions. The flow-based classification method (such as flow bytes per second, duration per flow, etc.) was used to characterize Non-VPN and VPN traffic using only time-related features. Each row in the dataset has 24 features, including class labels, and has more than eighteen thousand records.
The ISCX dataset has fourteen application labels (or classes), where seven of them were collected using VPN sessions, and the same seven were collected using regular Non-VPN encrypted sessions, as shown in Table \ref{tab1}. We use in our experiments 15,005 records for training and 3,752 for testing. The ISCX VPN-NonVPN dataset is a combination of both VPN and Non-VPN encrypted sessions.

\begin{table}[h]
    \normalsize
    \begin{center}
    \caption{List of ISCX Encrypted Traffic}
    \label{tab1}
        \begin{tabular}{||l|l||}
        \hline
            \multicolumn{1}{||c|}{\textbf{Traffic Type}} & \multicolumn{1}{|c||}{\textbf{Content}} \\ \hline \hline
            Browsing & Firefox, Chrome\\ \hline
            Chat & ICQ, AIM, Skype, Facebook, Hangouts\\ \hline
            Streaming & Vimeo, Youtube, Netflix, Spotify\\ \hline
            Mail & Email, Gmail ( SMPT, POP3,IMAP ) \\ \hline
            VoIP & Facebook, Skype, Hangouts, VoipBuster\\ \hline
            P2P & uTorrent, Bittorrent \\ \hline
            File Transfer & Skype, FTPS, SFTP \\ \hline
        \end{tabular}
    \end{center}
\end{table}

NIMS Dataset \cite{alshammari2011can} \cite{alshammari2008investigating} \cite{alshammari2007flow} is a captured network traffic set consisting of packets collected internally at the authors’ research testbed network. The dataset is available in two data formats, ARFF format and CSV format. The dataset is labeled. Different network scenarios are emulated using one or more computers to capture the resulting traffic. SSH connections are generated by connecting a client computer to four SSH servers outside the testbed via the Internet. Each row in the dataset has 23 features, including class labels, and has more than fourteen thousand records. It has six classes of encrypted SSH labels, as shown in Table \ref{tab2}. In our experiments, we use 11,408 records for training and 2,936 for testing.

\subsection{Experiments I: ML vs DL in Adversarial-Free Environment}
In the first set of experiments I, our objective is to build a highly accurate model on ISCX VPN-NonVPN and NIMS for each architecture: C4.5, KNN, ANN, CNN, and RNN. We extracted the same five features after applying the feature selection algorithm from our preprocessed dataset to train our models. We also normalized the dataset to change the values of the numeric features that have a different range to increase classifier performance and accuracy using StandardScaler \cite{brownlee2016machine}. It helps to scale the data within a range to avoid building incorrect models while training or executing data analysis. After that, we need to reshape the train and test input for CNN and RNN models into three dimensions (3D) using reshape() function \cite{Brownlee}. In our experiment, we reshape the input for the ISCX VPN-NonVPN dataset into a matrix of size (15005 x 5 x 1) for the training set and size of (3752 x 5 x 1) for the testing set. For the NIMS dataset, we reshaped the input into a matrix of size (11408 x 5 x 1) for the training set and size (2936 x 5 x 1) for the testing set. We split the dataset into training and validation set to evaluate the performance of the models with 80\% and 20\%, respectively. In Experiment I, we set up the hyper-parameters for deep learning models as the following: batch size= 64 sample, learning rate $\rho = 0.01$, and 20 epochs. Then, we test our trained model prediction to comprehensively compare the performance of machine learning and deep learning for encrypted multi-class classification in an adversarial-free environment.

\subsection{Experiments II: Resilience to Adversarial Attacks/Samples}
In the second set of experiments II, our objective is to repeat Experiment I using the same algorithms and the same preprocessing steps but applying ZOO, PGD, and DeepFool evasion attacks to generate adversarial samples. Thus, we can test and evaluate each model’s resilience against adversarial attacks and compare the output with the results generated from using a clean dataset in Experiment I.

\begin{table}
    \normalsize
    \begin{center}
    \caption{List of NIMS Encrypted Traffic}
    \label{tab2}
        \begin{tabular}{||l||}
        \hline
            \multicolumn{1}{||c||}{\textbf{Traffic Type}} \\ \hline\hline
            Local Tunneling \\ \hline
            Remote Tunneling \\ \hline
            SCP \\ \hline
            SFTP \\ \hline
            X11 \\ \hline
            Shell \\ \hline
        \end{tabular}
    \end{center}
\end{table}

\subsection{Feature Selection}
Feature selection is a crucial part of machine learning to reduce data dimensionality, and extensive research was carried out for a reliable feature selection method \cite{chandrashekar2014survey}. Filter method feature selection uses statistical measures to correlate and evaluate the relationship between the input variable and the target variable to filter and choose the better input variables used in the model.

We use a multilabel classification technique called mutual information \cite{kraskov2004estimating} to achieve high accuracy and better performance. Mutual information is a measure of statistical independence with two main properties: it can measure any relationship between random variables, including nonlinear relationships, and is invariant under transformations in the feature space invertible and differentiable \cite{vergara2014review}.
We train the C4.5, KNN, ANN, CNN, and RNN using mutual information for ISCX VPN-NonVPN and NIMS. The Experiment was executed multiple times for each algorithm with different batches and epoch values to obtain the best accuracy. After various tries and observations, we improve our model performance by selecting the same top five features for each dataset listed in Table \ref{tab3}.

\subsection{Evaluation Criteria}
We present the evaluation metrics we use in all experimental phases. To evaluate the model performance, various metrics can be used, including model Precision Rate (PR), Recall Rate (RC), and F1-Score (F1), which are commonly used in the literature.

Now, we give an example of what we mean by TP, FP, and FN as commonly used in the literature. If we are talking about Browsing, TP is the number of instances correctly classified as Browsing. FP is the number of instances incorrectly classified as Browsing. And, FN will be the case where Browsing be incorrectly classified as something else (e.g. Chat) \cite{wang2018deep}.

\begin{table}
    \normalsize
    \begin{center}
  \caption{The top 5 features selected from ISCX VPN-NonVPN and NIMS Datasets}\label{tab3}
  \begin{tabular}{|| c | l | c | l ||}
    \hline
    \textbf{ID} & \multicolumn{1}{||c|}{\textbf{ISCX}} & \textbf{ID} & \multicolumn{1}{c||}{\textbf{NIMS}}\\
    \hline\hline
     0 & duration & 16 & duration\\
     5 & max-fiat & 10 & max-fiat\\
     6 & max-biat & 14 & max-biat\\
     7 & mean-fiat & 9 & mean-fiat\\
     8 & mean-biat & 13 & mean-biat\\
    \hline
  \end{tabular}
    \end{center}
\end{table}

\begin{itemize}

\item \textbf{Precision}: The ratio of real positive samples over the total predicted positive samples \cite{wang2018deep}.\\

\begin{center}
\normalsize
$
Precision = \frac{TP}{TP + FP}
$
\end{center}

\item \textbf{Recall}: It is also called True Positive Rate (TPR) or detection rate. It is the total number of True Positives (TP) among all actual positive samples \cite{wang2018deep}.\\

\begin{center}
\normalsize
$
Recall = \frac{TP}{TP + FN}
$
\end{center}

\item \textbf{F1-Score}: It is a measure of a model’s accuracy on a dataset. It evaluates the binary classification systems, which classify examples as positive or negative. The F1-score is a way of combining the Precision and Recall of the model and defined as the harmonic mean of the model’s precision and recall \cite{wang2018deep}.

\begin{center}
\small
$
F1Score = 2 * \frac{Precision * Recall}{Precision + Recall}
$
\end{center}
\end{itemize}

\section{Results}
In this research, two experiments were designed using ISCX VPN-NonVPN and NIMS datasets to study the performance of machine learning and deep learning models for encrypted multi-class classification. This section elaborates on the experimental results. Instead of presenting each application (label) result, we took the average application results for every model. We use an average precision and recall to calculate F1-Score to evaluate the performance of ML/DL models.

\subsection{Results of Experiment I}
\begin{figure}[!ht]
  \centering
  \includegraphics[width=3.5in]{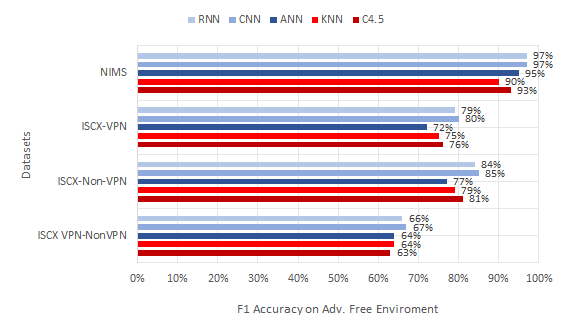}
  \caption{Evaluation Metric Results (\%) of five models in adversarial-free environment. DL algorithms are shown in different shades of Blue. ML algorithms are shown in different shades of Red.}\label{fig:1}
\end{figure}

We train two machine learning, i.e., C4.5 and KNN, and three deep learning models, i.e., ANN, CNN, and RNN, on two adversarial-free datasets. Fig. \ref{fig:1} shows the performance metrics of all models in the adversarial-free environment using the top five selected features in Table \ref{tab3}.

We can observe that deep learning models demonstrate better performance, on average, in the adversarial-free environment in both datasets: ISCX VPN-NonVPN and NIMS, in comparison to machine learning models.

\subsection{Results of Experiment II}
\begin{figure}[!ht]
  \centering
  \includegraphics[width=3.5in]{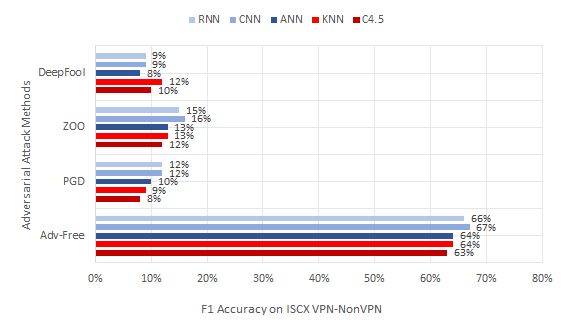}
  \caption{Effect of Adversarial Samples generated on five models trained by Adversarial-free ISCX VPN-NonVPN Combined. DL algorithms are shown in different shades of Blue. ML algorithms are shown in different shades of Red.}\label{fig:2}
\end{figure}

\begin{figure}[!ht]
  \centering
  \includegraphics[width=3.5in]{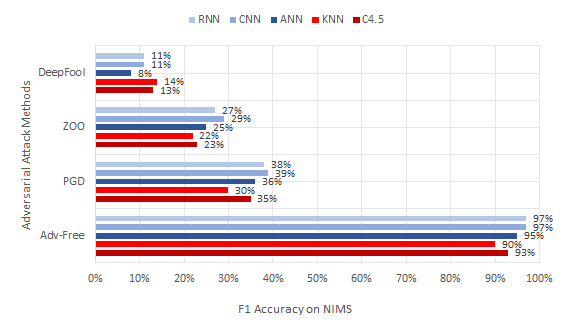}
  \caption{Effect of Adversarial Samples generated on five models trained by Adversarial-free NIMS dataset. DL algorithms are shown in different shades of Blue. ML algorithms are shown in different shades of Red.}\label{fig:3}
\end{figure}

We evaluate the resilience of the trained models against the crafted adversarial samples (using ZOO, PGD, and DeepFool evasion attacks). In this experiment, we build the learning models with different architecture, and we preprocess the benchmark datasets: ISCX VPN-NonVPN and NIMS using mutual information. Fig. \ref{fig:2} and \ref{fig:3} compare the effect of the adversarial attack methods against the learning algorithms. From Fig. \ref{fig:2} and \ref{fig:3}, we can see that machine learning and deep learning model accuracy were degraded significantly by the adversarial samples generated from the two datasets using the ZOO, PGD, and DeepFool evasion attacks.

In the ISCX dataset in Fig. \ref{fig:2}, we can conclude that, on average, deep learning models were more resilient to ZOO and PGD attacks compared to machine learning models. We also find that the resilience of deep learning was lower against the DeepFool attack than other attacks.

In NIMS dataset, deep learning performance shows more resilience against ZOO and PGD adversarial samples compared to machine learning, as shown in Fig. \ref{fig:3}, unlike DeepFool attack, where deep learning was less resilient to this type of attack compared to machine learning. We can also conclude that the DeepFool attack method has the highest effectiveness on NIMS dataset, similar to the ISCX VPN-NonVPN dataset.

We did another experiment to evaluate and test our models against adversarial samples by dividing the ISCX dataset into two parts: Non-VPN and VPN. Each dataset has more than nine thousand records, with seven application labels each. We used the same five features applied in other experiments from Table \ref{tab3}. We use 7,172 records for training and 1,793 for testing in the Non-VPN experiment, 7,835 records for training and 1,958 for testing in the VPN experiment.



\begin{figure}[!ht]
 \centering
 
\subfloat[ISCX-VPN]{%
  \includegraphics[width=3.5in]{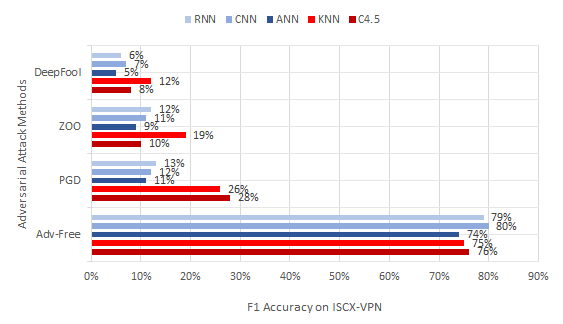}
  }

\subfloat[ISCX-NonVPN]{%
  \includegraphics[width=3.5in]{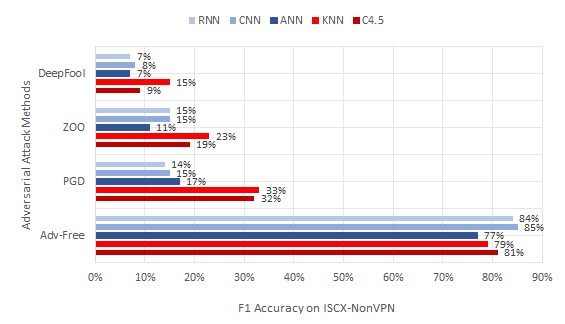}
  }
  
  \caption{Effect of Adversarial Samples generated on five models trained by Adversarial-free ISCX VPN and NonVPN Individually. DL algorithms are shown in different shades of Blue. ML algorithms are shown in different shades of Red.}\label{fig:4}
\end{figure}

From Fig. \ref{fig:4}, we can see, on average, that machine learning was more resilient to the three adversarial attacks compared to deep learning models, which rely on big datasets to avoid overfitting. Overfitting is a general problem when using deep learning algorithms \cite{shorten2019survey} \cite{lateh2017handling}. The amount of data needed to train a DL model is problem-dependent.

\section{Conclusion}
In this paper, we performed a comparison between machine learning and deep learning models in classifying encrypted traffic. We used different ML/DL models and observed that in an adversarial-free environment, on average, DL models perform better than ML models in terms of classification. Whereas, in the presence of an adversarial attack, the resilience of ML/DL would depend on the type of attack. In our experiments, the DeepFool evasion attack had the most impact on both models. Lastly, we conclude that DL models, on average, perform better than ML models against the tested adversarial evasion attacks in our experiments. 

\section*{Acknowledgment}
This work was supported by the Natural Sciences and Engineering Research Council of Canada (NSERC) through the NSERC Discovery Grant program.

\bibliography{references}
\bibliographystyle{ieeetr}

\end{document}